%% file: 5590FrozenSUSYpaper.tex
\font\caps=cmcsc10 at 12pt
\font\eightrm=cmr8 at 8pt
\documentclass{oxarticle}
\usepackage{tikz}

\usetikzlibrary{backgrounds,fit,decorations.pathreplacing,calc}

\usepackage{amsmath, amssymb}
\input{Johnmacros25}

\newcounter{orange} 
\newcounter{apple} 
\addtocounter{apple}{1}

\newcounter{grape} 
\addtocounter{grape}{1}
\newcommand{\numberhere}{5590}
\newcommand{\articlenumber}{\numberhere\; FrozenSUSYpaper.tex}

\proofmodefalse
\usepackage{color} 

\begin{document}


\renewcommand{\thefootnote}{\fnsymbol{footnote}}

 
 \begin{center}

{ \Huge  Frozen SUSY \\
with Susyons as the Dark Matter\\[1cm]}

\vspace*{.1in}

{\caps John A. Dixon\footnote{jadixg@gmail.com or john.dixon@ucalgary.ca}\\ University of Calgary,\\
Calgary, Alberta, Canada} \\[.5cm] 

\end{center}

 \begin{center}
 Abstract 
\end{center}

Frozen SUSY is the  maximally suppressed Supersymmetric SU(5)  Grand Unified Theory coupled to Supergravity. In Frozen SUSY, there is only one extra particle in addition to those that appear in the usual non-supersymmetric SU(5)  Grand Unified Theory coupled to gravity. Frozen SUSY also restricts and improves the mass predictions, and the cosmological constant (at tree level). As a result, it uses 4 parameters to generate 13 reasonable predicted masses.  The one extra particle is an extremely massive gravitino, which we call the Susyon.  In Frozen SUSY, the Susyon is stable and it interacts purely through gravity. The Susyon might be a viable candidate for dark matter.

 \refstepcounter{orange}
{\bf \theorange}.\;
{\bf \sS\ was introduced in \cite{Dixon:2018dzx,Dixon:2017eej}.}
The details of the basic mechanism for \sS\ were discussed in \ci{Dixon:2018dzx}. Then the application of that mechanism to the SUSY SU(5)  Grand Unified Theory coupled to Supergravity was discussed in \ci{Dixon:2017eej}.  
This paper uses the results of those papers, and discusses the maximally suppressed version of \ci{Dixon:2017eej}.  
This we call Frozen SUSY, and it has a number of useful features that were not considered in \ci{Dixon:2017eej}.

 \refstepcounter{orange}
{\bf \theorange}.\;
{\bf No superpartners have been found in the many experiments that have looked for them \ci{buchmueller,ccnonzero,darkenergy,Tanabashi:2018oca}.}   The methods of \sS\ allow us to write down a new theory, which we call `Frozen SUSY', in which all possible superpartners (except the gravitino) are `frozen'. Frozen SUSY is governed by a Master Equation and BRST cohomology and so it is a `genuine theory', rather than an `effective theory', 
a distinction explained in \ci{Dixon:2018dzx}.

 \refstepcounter{orange}
{\bf \theorange}.\;
{\bf Because Frozen SUSY does not have any of the usual superpartners for the observed particles,  it does not need spontaneous breaking of SUSY, or an invisible sector.} In the Gauge/Higgs sector there are only 4 parameters, but the theory predicts 13 masses from those (plus the susyon mass), and the 13 masses appear to be physically reasonable.  These can be found below in paragraph \ref{massesfromfs}. This economy arises because SUSY is still present in Frozen SUSY, through the Master Equation, and the resulting BRST cohomology.

 \refstepcounter{orange}
{\bf \theorange}.\;
{\bf  The Susyon:} This   gravitino in Frozen SUSY is stable with a known mass, predicted by the theory. Essentially, its mass comes from the Higgs mass and the Planck mass through a see-saw mechanism\footnote{This is buried in a complicated calculation in \ci{Dixon:2017eej}, and it needs a careful explanation.}.  Moreover, its only interaction with other matter is through gravitons. This gravitino will be called the Susyon in this paper.  The reason for the stability, and the restricted interactions, of the Susyon, is that all the superpartners are replaced by Zinn Sources using the ideas of \sS.   This paper discusses these features  below.

 \refstepcounter{orange}
{\bf \theorange}.\;
{\bf  \fS\ predicts the  Susyon:}  The reason for this prediction is that we cannot remove the gravitino from \fS. This is because the suppression trick of \sS\ is not applicable for gauge particles.  To see why this is true, we will start by describing how and why it works for scalars and spin $\fr{1}{2}$ fermions.

 \refstepcounter{orange}
{\bf \theorange}.\;
{\bf  \sS\ and Scalar Particles:}   We can focus on the following simple terms in the matter part of the Master Equation for the 
Action $\cA$ of a typical theory with SUSY and Gauge Transformations:
\be
\cM[\cA] = 
\int d^4 x \lt \{ \cdots +
\fr{\d \cA}{\d H^i}
\fr{\d \cA}{\d \G_i}
+
\fr{\d \cA}{\d \y^{i\a}}
\fr{\d \cA}{\d Y_{i\a}}
+
\fr{\d \cA}{\d \lam^{a\a}}
\fr{\d \cA}{\d M^a_{\a}}
+\cdots
\rt \}
\la{Mmatter}\ee
In the above expression $ H^i$ is a Complex   Grassmann even Scalar Field, $ \y^{i\a}$ is a   Grassmann odd Spin $\fr{1}{2}$ field, $ \G_i$ is a  Grassmann odd Zinn Source for the variation of the Scalar Field $ H^i$ and
$Y_{i \a}$ is a    Grassmann even Zinn Source for the variation of the Spin $\fr{1}{2}$ field $ \y^{i\a}$.  The index $i$ labels different fields, and the index $\a$ is a complex Weyl spinor index $\a=1,2$. We have also added gauginos $\lam^{a\a}$ and a Zinn Source for them $M^a_{\a}$, where the index $a$ is an index for the gauge group in its adjoint representation. 

 \refstepcounter{orange}
{\bf \theorange}.\;
{\bf Now let us look at some simple terms in the action:}
\be
\cA_{\rm Scalar\;Fields} = 
\int d^4 x \lt \{
H^i \Box \oH_i 
+ \G_i \lt (
C^{\a} \y_{\a}^i
+ iT^{a i}_{\;\;j} H^j \w^a
+ \cdots \rt ) 
+ \cdots
\rt\}
\la{Ascalarmatter}
\ee
 The terms $\G_i \d H^i=
\G_i \lt (
C^{\a} \y_{\a}^i
+i T^{a i}_{\;\;j} H^j \w^a
+ \cdots\rt ) $ contain the BRST variation $\d H^i$ of the field $H^i$. The BRST operator $\d$ is a  Grassmann odd, nilpotent, mapping. 
 We have not included interaction terms and gravitational variation terms here because they would not change our conclusions. In (\ref{Ascalarmatter}),
$ C^{\a}$ is a  Grassmann even Spin $\fr{1}{2}$ supersymmetry Ghost field,  
$ \w^a$ is a  Grassmann odd Spin 0 Faddeev-Popov Ghost field 
and $ T^{ai}_j$ are group representation matrices.

 \refstepcounter{orange}
{\bf \theorange}.\;
\la{scalarparagraph}
{\bf First we will examine  transformations that change the first part of the above equation (\ref{Mmatter}):}
\[
\cM[\cA] = 
\int d^4 x \lt \{
\fr{\d \cA}{\d H^i}
\fr{\d \cA}{\d \G_i}
+\cdots
\rt \}
\]
\be
\Ra
\cM_{\rm New}[\cA_{\rm New}] = 
\int d^4 x \lt \{
\fr{\d \cA_{\rm New}}{\d J^i}
\fr{\d \cA_{\rm New}}{\d \h_i}
+\cdots
\rt \}
\la{Mmattervar}
\ee
We are introducing new variables here.
$ H^i$ was a  Grassmann even  Scalar field and it gets replaced by $ J^i$, which is a new Grassmann even  Scalar Zinn Source  and $ \G_i$ was a Grassmann odd   Zinn Source 
and it gets replaced by $\h_i$, which  is a new Grassmann odd  Antighost Scalar Field. This sort of `Exchange Transformation' is discussed at length in the papers \cite{Dixon:2018dzx,Dixon:2017eej}.

 \refstepcounter{orange}
{\bf \theorange}.\;
\la{scalarreplaceparagraph}
{\bf
The expression  
(\ref{Ascalarmatter}), after the above change of variables, becomes:}
\[ \cA_{\rm New\; Theory\; From\; Scalars}[J, \h]=
\cA_{\rm Scalar\;Fields}[H\ra J, \G\ra \h,\cdots] =\]\be
\int d^4 x \lt \{
J^i \Box \oJ_i 
+ \h_i \lt (
C^{\a} \y_{\a}^i
+ T^{a i}_{\;\;j} J^j \w^a
+ \cdots
\rt ) 
+ \cdots\rt\}
\la{Anewfromscalars}
\ee

 \refstepcounter{orange}
{\bf \theorange}.\;
\la{awaywithkineticscalar}
{\bf   In the old path integral that defined the original theory,} we integrated over $H^i$, but not over $\G_i$.  This was because $H^i$ was a quantized scalar field, and $\G_i$ was just a `Zinn 
Source', originally introduced into these expressions by Zinn Justin 
\ci{poissonbrak,Becchi:1975nq,zinnbook,taylor,Zinnarticle,becchiarticles1,becchiarticles2,BV}.
The purpose of introducing the Zinn Sources was to allow us to write the Master Equation in a quadratic form like (\ref{Mmatter}).  This form is crucial to \sS, because it has some of the properties of a Poisson Bracket.  

 \refstepcounter{orange}
{\bf \theorange}.\;
\la{nopropforscalarghost}
{\bf   In the new path integral that will define the new theory with \sS,} we choose to integrate over the new antighost variable $\h_i$, and not over the new Zinn Source $J^i$.  Note that $J^i \Box \oJ_i $ in (\ref{Anewfromscalars}) 
is not quantized, even though it looks like a kinetic term.   This is essential   because $J$ is a Zinn  Source.    The new field $\h_i$ is  a quantized `antighost field'.  But $\h_i$ does not propagate, because it does not appear in any quadratic term in the action. However a term like
$ \h_i 
C^{\a} \y_{\a}^i$ in (\ref{Anewfromscalars}) is a perfectly acceptable trilinear interaction term in the new theory. It just does not get to do anything much in the quantum field theory, because $ \h_i $ does not appear in any quadratic term, so it cannot give rise to Feynman diagrams with propagators involving $\h_i$.

 \refstepcounter{orange}
{\bf \theorange}.\;{\bf So in some sense we have {\bf `frozen out'} the scalar $H^i$ degree of freedom,} while maintaining the SUSY algebra, and keeping the same old SUSY Master Equation, but with new names and new properties for the scalar field and its Zinn source.

 \refstepcounter{orange}
{\bf \theorange}.\;
{\bf   
Thus, as far as this part goes,} we had, for this piece of the original  \GUST:
\be
Z[j, \G, \cdots]= \P_{x,i} \int \d H^i(x) \cdots e^{i \lt \{ \cA +\int d^4 y j_i H^i + \cdots\rt \}
}
\la{actioncA}
\ee
where $\cA$ contains (\ref{Ascalarmatter}) and $j_i$ is a new source inserted to couple to $H^i$, and this piece becomes, for this part of the \SGUST:
\be
Z'[J, \z,\cdots]= \P_{x,i} \int \d \h_i (x) \cdots e^{i  \lt \{ \cA' +\int d^4 y \z^i \h_i  + \cdots\rt \}}
\la{actioncAp}
\ee
where $\cA'$ contains  (\ref{Anewfromscalars}),  and $\z^i$ is a new source inserted to couple to $\h_i$.

 \refstepcounter{orange}
{\bf \theorange}.\;
{\bf   
We get a new integral and a new generating functional $Z$.} 
There is a standard and simple formal derivation involving a Legendre transform and a connected generator of Green's functions here \ci{poissonbrak}\footnote{We will assume that nothing goes wrong with that argument in the present case.   But that is not very obvious even in the normal situation.  It requires a proof and a demonstration for this special and unusual kind of case.   I  expect that the result of that is the usual one.}.
 This means that the invariance is such that we will get: 
\be
\cM[\cG] = 
\int d^4 x \lt \{
\fr{\d \cG}{\d H^i}
\fr{\d \cG}{\d \G_i}
+\cdots
\rt \}
\ee
\be
\Ra
\cM_{\rm New}[\cG_{\rm New}] = 
\int d^4 x \lt \{
\fr{\d \cG_{\rm New}}{\d J^i}
\fr{\d \cG_{\rm New}}{\d \h_i}
+\cdots
\rt \}
\ee
where the $\cG$ are the 1PI generating functionals for the new and old theories.  This is certainly true for the tree level, where the 1PI generating functionals are exactly the action $\cA$ and $\cA'$ referred to in (\ref{actioncA}) and (\ref{actioncAp}), which arise in turn from  (\ref{Ascalarmatter}) and (\ref{Anewfromscalars}).  The case where we go beyond tree level needs examination from many points of view, and will not be attempted here.  It is an important next step however.

 \refstepcounter{orange}
{\bf \theorange}.\;
{\bf   Fermions Work in a Very Similar Way to the Scalars:} 
Here are the basic terms of the action, by analogy with (\ref{Ascalarmatter}): 
\be
\cA_{\rm Spinor\;Fields} = 
\int d^4 x \lt \{
\y^{i\a} \pa_{\a \dot \b} \oy_{i}^{ \dot \b}
+ Y_{i\a} \lt (
\pa^{\a \dot \b} H^i \oC_{\dot \b}
+   F_H^i C^{\a}
+ iT^{a i}_{\;\;j} \y^{j\a} \w^a+ \cdots \rt ) 
+ \cdots
\rt\}
\la{Aspinormatter}
\ee
 The terms $Y_{i\a} \d \y^{i\a} =
Y_{i\a} \lt (
\pa^{\a \dot \b} H^i \oC_{\dot \b}
+   F_H^i C^{\a}
+i T^{a i}_{\;\;j} \y^{j\a} \w^a + \cdots\rt ) 
$ contain the BRST variation $ \d \y^{i\a} $ of the field $\y^{i\a}  $. 
The new field here is $F_H^i$, which is the well known auxiliary $F$ field that goes with the scalar $H^i$.  
 Again we have not included interaction terms and  gravitational variation terms here because they would not change our conclusions.

 \refstepcounter{orange}
{\bf \theorange}.\;
{\bf Now consider the fermionic version of (\ref{Mmattervar}) above:}
\[
\cM[\cA] = 
\int d^4 x \lt \{
\fr{\d \cA}{\d \y^{\a}}
\fr{\d \cA}{\d Y_{\a}}
+\cdots
\rt \}
\]
\be
\Ra
\cM_{\rm New}[\cA_{\rm New}] = 
\int d^4 x \lt \{
\fr{\d \cA_{\rm New}}{\d \S^{\a}}
\fr{\d \cA_{\rm New}}{\d G_{\a}}
+\cdots
\rt \}
\ee
We are performing exactly the same exercise with the spinors as we did with the scalars above in paragraph \ref{scalarparagraph} above. We are again introducing new variables here.
$ \y^{i\a}$ was a  Grassmann odd Spinor field and it gets replaced by $\S^{i \a}$, which is a Grassmann odd New Spinor Zinn Source  and $ Y_{i \a}$ was a Grassmann even  Zinn Source 
and it gets replaced by $G_{i\a}$, which  is a Grassmann even  New  Antighost Spinor Field.

\refstepcounter{orange}
{\bf \theorange}.\;
\la{spinorreplaceparagraph}
{\bf
The expression  
(\ref{Aspinormatter}), after the above change of variables, becomes:}
\be
\cA_{\rm New\; Spinor\;Fields} = 
\int d^4 x \lt \{
\S^{i\a} \pa_{\a \dot \b} \ov \S_{i}^{ \dot \b}
+ G_{i\a} \lt (
\pa^{\a \dot \b} H^i \oC^{\dot \b}
+   F_H^i C^{\a}
+ iT^{a i}_{\;\;j} \S^{j\a} \w^a+ \cdots \rt ) 
+ \cdots
\rt\}
\la{newspinoract}
\ee

  \refstepcounter{orange}
{\bf \theorange}.\;
\la{nopropforspinorghost}
{\bf   In the new path integral that defines the new theory,} instead of integrating over $\y$, and not over $Y$ as in the old path integral, 
we integrate over $G$, and not over $\S$. 
Note that $\S^{i\a} \pa_{\a \dot \b} \ov \S_{i}^{ \dot \b}
$ is not quantized because $\S^{i\a}$ is a Source.  
  The new field $G$ is  a quantized `antighost spinor field'.
 But $G$ does not propagate, because it does not appear in any quadratic term in the action.   So in some sense we have now {\bf `frozen out'} the spinor $\y^{i \a}$ degree of freedom, while maintaining the SUSY algebra, and keeping the same old SUSY Master Equation, but with different names and properties--as to whether some quantities are  unquantized Zinn Sources or quantized Fields.  This is all completely analogous to paragraph \ref{awaywithkineticscalar} above which dealt with the scalar case.

  \refstepcounter{orange}
{\bf \theorange}.\;
{\bf   The auxiliary field $F_H^i $ needs special attention:}  In the case of 
(\ref{newspinoract}) where we are only changing the spinor field and its Zinn Source, we do not need to change the auxiliary field $F_H^i $.  But what if we consider the situation where we introduce the change 
in paragraph (\ref{scalarparagraph})?  Here we need to remember that the auxiliary fields can be integrated explicitly in the path integral, and the result is that one gets expressions that are polynomials in the other fields in the action from that.  For example we might have
\be
 F_H^i {\ov  F}_{H i}  + F_H^i \lt ( Y_{i\a}    C^{\a} + g_{ijk} H^j H^k \rt )+ h.c. \Ra - \lt ( Y_{i\a}    C^{\a} + g_{ijk} H^j H^k \rt ) 
 \lt ( \ov Y^i_{\dot\a}    \ov C^{\dot\a} + \ov g^{ilm} \oH_l \oH_m \rt )
 \ee
in the action.  The second form is the result of the integration of the first over $ F_H^i$ in the path integral--it gives the quadratic form shown as part of the new action after integration.

  \refstepcounter{orange}
{\bf \theorange}.\;
{\bf  The point is that we can eliminate all the auxiliary fields in this way, and the resulting Master Equation still looks the same.}  So when we change the $H^i$ variable this gets taken care of automatically.  We do not need to worry about changing variables, or about having Zinn sources for, the auxiliary variables like  $F_H^i$.  The nilpotence coming from the Master Equation still works too after this integration, with the variation being supplied by the variation of the Zinn Sources from the Master Equation, as one can verify.

  \refstepcounter{orange}
{\bf \theorange}.\;
{\bf  What if there is spontaneous breaking of supersymmetry?}  In this case it follows that some auxiliary fields must have non-zero vacuum expectation values \ci{freepro,xerxes,haber,Weinberg2,Weinberg3,ferrarabook,superspace,WB, Buchbinder:1998qv,west,pran}.   This is a complication that could cause trouble.  But it is exactly the complication that we avoid with Frozen SUSY, at least at tree level, since we assume that supersymmetry is not spontaneously broken in the papers  \ci{Dixon:2018dzx} and \ci{Dixon:2017eej}.  Again, there are issues here that need to be looked at carefully at higher orders. Does the cosmological constant reappear at higher orders?  That might even be desirable given the experimental results on dark energy \ci{darkenergy,Tanabashi:2018oca}, and work such as that in 
\ci{xerxes,haber} might still apply in some sense at higher orders.

  \refstepcounter{orange}
{\bf \theorange}.\;
{\bf  The same story happens for gauginos $\lam^{a \a}$ as for chiral spinors given above.} The action for gauginos contains an auxiliary $D^a$ and the field strength $F^a_{\a \b}$. But these do not change the above arguments at all.

 \refstepcounter{orange}
{\bf \theorange}.\;
{\bf   However the story changes for gauge fields. They do not work the same way, because gauge fields have an inhomogeneous term in their variation.} For example, consider a vector gauge field: 
\ben
\item
 This would yield a quadratic term of the form
\be
\int d^4 x\;
\S^{\m a} \pa_{\m} \w^a
\Ra
\int d^4 x\;
\h^{\m a} \pa_{\m} \w^a
\la{propghost}
\ee
where we imagine taking a Zinn Source $\S^{\m a}$ and changing it to an antighost field $\h^{\m a}$, like we did above for the scalar example.  
\item
The problem is that this would 
 wreak havoc with the theory after the Exchange Transformation, because this antighost  $\h^{\m a} $ would propagate because of the quadratic term (\ref{propghost}) above. The argument in paragraphs 
\ref{nopropforscalarghost} and \ref{nopropforspinorghost} above 
 do not work in this case. \item
  So Suppressed SUSY can only affect the scalars and spin $\fr{1}{2}$ fields, including both chiral spinors and gauginos, but not the vector bosons, and not the gravitino nor the graviton.  These all have terms like the ones in 
  (\ref{propghost}). 
  \item
  This is why we end up with a massive gravitino in Frozen SUSY:  the Susyon.  But that is not a bad feature: it looks like it could account for the observed dark matter.
 \een

\refstepcounter{orange}
{\bf \theorange}.\;
{\bf Origin of the Mass of the SU(5) \sS\ Theory Gravitino:} From page  388 of \ci{freepro}, we see that the relevant terms for this are, from the first row of (18.6) and expression (18.7):
\be
e^{-1} {\cal L}_{1} = \fr{1}{2 \kappa^2} \ov \y_{\m} \g^{\m \r \s} \pa_{\r} \y_{\s}
\ee
plus, from (18.15):
\be
  {\cal L}_{ m, \fr{3}{2}} = \fr{1}{2} m_{\fr{3}{2}} \ov \y_{\m} P_{\rm R}
  \g^{\m \n} \y_{\n} +{\rm h.c.}
\ee
where from (18.16) of \ci{freepro} we have
\be
   m_{ \fr{3}{2}} = \kappa^2 e^{\kappa^2 \cK/2} W
\la{massofsusyon}
\ee
We note that the expression $P_L v$ in (19.1) of  \ci{freepro} is zero for \fS.  So indeed there is  no Goldstone fermion here and no spontaneous breaking of SUSY here.

However there is a huge gravitino mass from 
(\ref{massofsusyon}), even though the related cosmological constant is zero at tree level.  Note that the local supersymmetry invariance is still present in this action, because the Master Equation is still true by construction.  
There is work to be done to examine the BRST cohomology of this theory at one loop (and higher levels).  Even writing the Master 
Equation down in its full form is a major chore, and we do not propose to do it here.  But it needs to be done.

 \refstepcounter{orange}
{\bf \theorange}.\;
{\bf Details of the Mass of the SU(5) \sS\ Theory Gravitino:}  From the paper \ci{Dixon:2017eej} we have
\be
M_{\rm Gravitino} \equiv 
M_{\y}= 
1.64 M_{\rm SP} 
\ee
where
\be
 M_{\rm SP} 
 = 1.22 \times 10^{36} g_5^2 {\rm GeV/c^2}
; 1 {\rm GeV/c^2}= 1.78 \times 10^{-27} {\rm kg.}
 \eb
 M_{\rm Gravitino} \equiv 
M_{\y}= 
1.64  \times 1.22 \times 10^{36} g_5^2 
 \times 1.78 \times 10^{-27}
 {\rm kg.} 
 \eb
 \approx  10^{9}
 {\rm kg.} \approx  1 {\rm Megatonne} 
 \approx  8 \times  {\rm Mass\;of\;  a\; Royal\; Caribbean \;Cruiseship}
 \ee
 We call these Susyons. We have taken $g_5\ra \fr{1}{2}$ for no particularly good reason.  This  $g_5$ needs to be determined with the kind of renormalization group arguments that also explain $\sin \q_W$. These can be found in say \ci{Weinberg2,Weinberg3}. I do not know if some problems arise in this context, and they might, given that the masses here are determined from the scalar potential, whereas in the usual treatment they are determined by the desire to have the three gauge couplings converge to one.

 \refstepcounter{orange}
{\bf \theorange}.\;
{\bf Detailed Reasons for Stability of the Minimal SU(5) \sS\ Theory Gravitino:}  
To see that the Susyon is stable, we need to examine the terms in the action that link the Gravitino $\y$ to other fermions. We shall see that none of these give rise to a decay mode for the Susyon. For the same reason, this analysis shows that the Susyon, in this Frozen SUSY theory, does not interact with anything other than gravity. From page 388 of \ci{freepro}, we see that these terms are of the following kinds:
\ben
\item
The term with only one Gravitino $\y$, linked to only one spin $\fr{1}{2}$ gaugino $\lambda$ (plus bosons). Here is this  term\footnote{We label these various terms with subscripts as in $ {\cal L}_{,5}$ }:
\be
e^{-1} {\cal L}_{,5}
=
 \fr{1}{8 } {\rm Re\; f}_{AB} {\ov \y}_{\m} \g^{ab} \lt ( F^A_{ab} + {\widehat F}^A_{ab} \rt )  \g^{\m} \lambda^B 
\ee
Since we have transformed all gauginos $\lambda$ to Zinn sources, this term becomes a Zinn Source term and so it cannot contribute to $\y$ decay\footnote{The supercovariant gauge curvature ${\widehat F}^A_{ab}$ is given in (18.14) of  \ci{freepro}.  It also contains a gaugino, which is also changed to a Zinn Source.}. 
 
\item
The term with only one Gravitino $\y$, linked to only one spin $\fr{1}{2}$ chiral multiplet fermion $\c$ (plus bosons). Here is this term:
\be
e^{-1} {\cal L}_{,6}
=
 \fr{1}{\sqrt{2} } {\rm g}_{\a {\ov \b}} {\ov \y}_{\m} {\hat \pas} {\ov z}^{\ov \b} \g^{\m} \c^{\a} +{\rm h.c.}
\la{L6terms}
\ee
Since we have transformed
\ben
\item
 all chiral scalars $z^{\a}$ in the chiral Matter (Quark and Lepton) multiplets to Zinn sources, and
 \item
 all chiral spinors $\c^{\a}$ in the chiral Higgs multiplets  to Zinn sources;
\een
  every term in (\ref{L6terms})
 becomes a Zinn Source term and so it cannot contribute to $\y$ decay. 
We note that the metric  ${\rm g}_{\a {\ov \b}}$ in (\ref{L6terms}) connects Squarks to Quarks, Sleptons to Leptons, and Higgsinos to Higgs scalars.

\item
The term with only one Gravitino, linked to three spin $\fr{1}{2}$ fermions (plus bosons). 
\be
{\cal L}_{{\rm 4 f}, 1}
=
- \fr{1}{4 \sqrt{2}} f_{AB\a} \oy .\g \c^{\a} {\ov \lambda}^A P_L \lambda^B 
\ee
Since we have transformed all gauginos $\lambda$ to Zinn sources, this term also becomes a source term and it cannot contribute to $\y$ decay. 
 
\item
The `mixed term' with only one Gravitino, linked to one spin $\fr{1}{2}$ gaugino. 
\be
{\cal L}_{{\rm mix}, 1}
=
\oy .\g  \fr{1}{2} i P_L \lambda^A {\cal P}_A 
\ee
Since we have transformed all gauginos $\lambda$ to Zinn sources, this term also becomes a source term and it cannot contribute to $\y$ decay. 
 
\item
The other `mixed term' with only one Gravitino, linked to one spin $\fr{1}{2}$ chiral multiplet fermion: 
\be
{\cal L}_{{\rm mix}, 2}
=
\oy .\g  \fr{1}{\sqrt{2}} \c^{\a} e^{\kappa^2 {\cal K}/2} \na_{\a} W \; + 
{\rm h.c.}
\ee
This last term contains terms that are chiral matter fermions multiplied by chiral matter scalars, which are Zinn  Sources, and it also contains Higgs fermions multiplied by Higgsino scalars.  The Higgs fermions have been also changed to Zinn Sources.  So all terms of this kind are Zinn Source terms and they cannot contribute to $\y$ decay.
 
 \item The fact that all these terms are really Zinn terms also means that there is no coupling between the Susyon and any other matter in the action, except the gravitons. This is a suggestive feature that might make the Susyon into a good candidate for the dark matter.
\een

 \refstepcounter{orange}
{\bf \theorange}.\;
{\bf  How many Susyons do we need to make up the missing mass of the universe, assuming that these are the only dark matter?} The estimated local dark matter density \ci{Tanabashi:2018oca} is
\be
\rho_{\rm DM} \approx .4 \;
{\rm GeV/cm^3} 
\ee
and this means that we have, for each gravitino, a volume of approximately 
\be
V_{\rm Gravitino} 
\approx
\fr{ 10^{36} {\rm GeV}}{
.4 \; {\rm GeV/cm^3} }
\approx
 2 \times 10^{36} {\rm cm}^3
\approx
{ \{ 10^{12} {\rm cm} \} }^3
\eb
\approx
{ \{ 10^{10} {\rm m} \} }^3
\approx
{ \{ 10^{7} {\rm km} \} }^3
\approx
{ \{ 10 \; {\rm million}\; {\rm km} \} }^3
\ee
The distance from the earth to the sun is 93 million miles, or about 150 million km.  So this is about $\fr{1}{15}$ times the distance to the sun.

 \refstepcounter{orange}
{\bf \theorange}.\;
{\bf  Their number density if they are the dark matter?}  There are about $15^3=3375$ Susyons in the volume contained in a sphere whose radius is the distance from the sun to the earth. 
 
 \refstepcounter{orange}
{\bf \theorange}.\;
{\bf  Their average velocity if they are the dark matter?}  Presumably this is of the order of the escape velocity from our galaxy. This implies that one susyon hits the earth about every century. Since the sun's area as a target is about $10^4$ times that of earth, we could expect that the sun gets hit by a susyon about one hundred times a year.

 \refstepcounter{orange}
{\bf \theorange}.\;
{\bf  The number of Susyons passing through a typical room on earth?} The probability here is something like once every million universe lifetimes. 

   \refstepcounter{orange}
{\bf \theorange}.\;
{\bf  Cross Section and Energy and Parameters of a Collision between a Susyon and some matter on earth?} How hard is this computation?  I do not know.  It is very complicated for sure. Also there is the question what happens in this theory beyond tree level.  That is not a simple question, since the theory is not renormalizable, and although string theory is obviously relevant, as it must be for any supergravity theory, it is not at all clear how to implement that with these Exchange Transformations.
 
 \refstepcounter{orange}
{\bf \theorange}.\;
{\bf  Simple First Try:} However at tree level, this is probably a feasible calculation.  The Susyon interacts only through gravity.  Can one assume that this can be done classically and non-relativistically?  Then it is rather like a combination of a simple matter of bending of a path in a gravitational field, with some sort of kinetic transport non-equilibrium calculation of the heating of the medium, and corresponding loss of momentum and energy for the Susyon.  The Susyon will only affect nearby particles, since gravity is so weak, even for a particle as massive as a Susyon. There is an interesting calculation to do here.  Does passage through a galaxy, and stars and dust and gas in that galaxy, slow down Susyons at the right rate to make a sensible story for dark matter?  

 \refstepcounter{orange}
{\bf \theorange}.\;
{\bf  If a Susyon goes through the earth near us, would we notice?}  Such an event would be extremely rare, but even when it happens, it seems likely that the effect would be small, because gravity is so weak, and the Susyon is just an elementary particle, even though it has a large mass.  If it interacted with other matter directly that might make a large difference.   But these questions need some serious work.

 \refstepcounter{orange}
{\bf \theorange}.\;
{\bf Schwarzlength and Broglielength:} We can define two kinds of length associated with any mass $m$:
\be
{\rm Schwarzlength}_{\rm m}=  \fr{G m }{c^2}
\ee
\be
{\rm Broglielength}_{\rm m} =  \fr{\hbar }{c m}
\ee
For example for a proton:
\be
{\rm Schwarzlength}_{\rm Proton} =  1.2 \times 10^{-52} {\rm cm.}
\ee
\be
{\rm Broglielength}_{\rm Proton} = 2 \times 10^{-14} {\rm cm.}
\ee
and, for example, for the earth:
\be
{\rm Schwarzlength}_{\rm Earth} =  .442 \; {\rm cm.}
\ee
\be
{\rm Broglielength}_{\rm Earth} = 5.8 \times 10^{-66} {\rm cm.}
\ee
Note that
\be
\fr{\rm Schwarzlength} {\rm Broglielength} =  
\fr{G m^2 }{c \hbar}
\ee

 \refstepcounter{orange}
{\bf \theorange}.\;
{\bf  The length at which the one becomes larger than the other is at:}
\be
\fr{\rm Schwarzlength} {\rm Broglielength} =  1=
\fr{G m^2 }{c \hbar}
\ee
and this happens at the Planck mass:
\be
m_{\rm Planck} = \sqrt{\fr{c \hbar}{G}} = 2.17316 \times 10^{-5} {\rm gm.}
\ee
and the Plancklength:
\be
 {\rm Schwarzlength}_{\rm Planck} =  {\rm Broglielength}_{\rm Planck} = 
  1.6 \times 10^{-33} {\rm cm.}
  \ee

\refstepcounter{orange}
{\bf \theorange}.\;
{\bf Planck Mass and Length and the Susyon:}
So for masses greater than the Planckmass the Schwarzlength is greater than the Broglielength.   Now the Susyon with mass $8.9 \times 10^{11} {\rm gm.} $ has its $ \rm Schwarzlength = 6.6 \times 10^{-17} {\rm cm}$,  much greater than its   $ \rm Broglielength = 3.9 \times 10^{-50} {\rm cm.}$ Since it is an elementary particle, one wonders whether this means it has a tendency to be a black hole, or something like that.  Of course, classical reasoning is not relevant here, presumably. But what {\em is} relevant?

\refstepcounter{orange}
{\bf \theorange}.\;
{\bf The Gauge/Higgs Gravitino Sector of Frozen SUSY:} This was worked out in  \ci{Dixon:2017eej}  with the aid of some computer code included there\footnote{To bring that code up to date for the current software as of this writing, the extra line  $\rm \$Assumptions=f>0$ is needed. The only difference from that paper is that Frozen SUSY assumes that all the Higgsinos, Squarks, Sleptons and Gauginos are suppressed. This makes no difference to these mass results.}.

\refstepcounter{orange}
{\bf \theorange}.\;
{\bf There are only four parameters $(g_1,g_2,g_5,M_P =\fr{1}{\kappa})$ for the Gauge/Higgs Gravitino sector,} and one gets  13  boson masses (the graviton, photon, gluons, Higgs, Z, W, X, Y and five very heavy extra Higgs), plus the Susyon mass from them.

\refstepcounter{orange}
{\bf \theorange}.\;
{\bf The superpotential has the form:}
\be
 W =
 e^{\fr{-1}{4}\k^2 \lt (2 H_{L}^i H_{Ri} + \fr{1}{2}\Tr S^2 
  \rt )} 
M_P^{\fr{3}{2}}  \sqrt{g_1 \Tr( S S S) - g_2 H_L^i S_i^{\;j}  H_{Rj}  }   
    \ee
where these scalar fields are the $5, \ov 5$ and complex $24$ of SU(5).
We could convert these dimensionless parameters $g_1,g_2$ to masses using 
\be
M_1 = -{g_{1}}^{\fr{1}{3}} M_P
;\;M_2 = {g_{2}}^{\fr{1}{3}} M_P
\ee
\refstepcounter{orange}
{\bf \theorange}.\;
{\bf We defined the following `Hierarchy \; Parameters', and the Bosonic Masses of W, Z and H required these values for $g_1, g_2$:} 
\be
\begin{array}{|c|c|}
 \hline
\multicolumn{2}{|c|}{\rm Tiny \; Parameters }\\
 \hline
  f g_5  =  M_W/M_P  &h = M_H/M_P  \\
 \hline
     3.35 \times 10^{-17}  g_5 &5.2 \times 10^{-17}   \\
\hline
\end{array}
\;
\begin{array}{|c|c|}
 \hline
\multicolumn{2}{|c|}{\rm The \; Superpotential\;Parameters }\\
 \hline
 g_1 &r  = {g_2}/{g_1}   \\
 \hline
 -5.81 \times 10^{34} g_5^4
  &   -2- 2 f^2/9  \\
\hline
\end{array}
\ee

\refstepcounter{orange}
{\bf \theorange}.\;
{ \bf We defined:} the Planck and SuperPlanck Masses
and the masses required the following Vacuum Expectation Values (VEVs):
\be
\begin{array}{|c|c|}
\hline
\multicolumn{2}{|c|}{\rm Mass\; Names\; (GeV) }\\
 \hline
  M_P & M_{\rm SP} \\
 \hline
2.4 \times 10^{18} &  1.22\times 10^{36} g_5^2\ \\
\hline
\end{array}
\;
 \begin{array}{|c|c|c| }
\hline
\multicolumn{3}{|c|}{\rm VEVs\; (GeV)} \\
 \hline
M_{\text{K9}} & M_{\text{K1}}  &M_{\text{K2}}    \\
 \hline
 -f M_{P}& -\fr{M_P f^2}{6
   \sqrt{10}}& -\sqrt{6}
   M_P
    \\
\hline
\end{array}
\ee
\refstepcounter{orange}
{\bf \theorange}.\;
\la{massesfromfs}
{\bf Then we arrived at the following 11 masses (plus the zero mass photon, gluons, and graviton and the zero value for the cosmological constant):} 
\be
 \begin{array}{|c|c|c|}
 \hline
\multicolumn{3}{|c|}{\rm Electroweak\; Masses\; \; Units\; of\;M_{\rm P}} 
\\
 \hline
{\rm M}_Z  = \fr{8 }{5}f g_5  M_P &{\rm M}_W =f g_5 M_P&{\rm M}_H  =h M_P   \\
\hline
\end{array}
\eb
\begin{array}{|c|c|}
\hline
\multicolumn{2}{|c|}{\rm Heavy \; Vector\;Bosons:  \; Units\; of\;M_{\rm P} }\\
 \hline
{\rm X}   &{\rm Y}     \\
\hline
 2\sqrt{10}  g_5 M_P &
2 \sqrt{10}  g_5 M_P \\
\hline
\end{array}
\eb
\begin{array}{|c|c|c|c|c|c|}
\hline
\multicolumn{6}{|c|}{\rm Super\; Planck\; Masses: \; Units\; of\;M_{\rm SP}}\\
 \hline
{\rm H_{\rm Oct}} &{\rm H_{Trip} }&{\rm H^+}&{\rm H}_2  &{\rm H}_3   
&{\rm Gravitino=Susyon}\\
\hline
2.05  &0.68   &   2.05  &  .81   &2.05   &  1.64 
\\
\hline
\end{array}
\ee
 \refstepcounter{orange}
{\bf \theorange}.\;
{\bf Conclusion:} This \fS\ theory predicts a Susyon that is terrifically heavy and presumably terrifically hard to observe.  But it also predicts a number of other things, and it raises a number of issues:
\ben
\item
The mass of the $X,Y$ vector bosons and the lifetime of the Proton are related to the masses of the Higgs, Z and W, and to the Planck mass. The heavier masses for the $X,Y$  and the five extremely heavy Higgs multiplets seem to be an improvement over the old models in \ci{ross, GUT}. 
\item
The theory allows for the cosmological constant to be naturally zero at tree level, in contrast to theories where there is spontaneous breaking of SUSY.
\item
The theory is very close to the observed standard model, and it predicts that there will be little more to discover, as far as new particles are concerned, short of super-Planck masses.  
\item The theory has a natural way of accounting for gauge symmetry breaking in the usual pattern $SU(5) \ra SU(3) \times SU(2) \times U(1)\ra SU(3) \times U(1)$. 
\item
There is no need for an invisible sector, or a messenger sector, or lots of effective coupling parameters.  These are familiar from the SSM \ci{xerxes,haber}, where they arise from the hypothesis of spontaneous supersymmetry breaking. 
\item
In fact, the present theory has only four parameters for the gauge/Higgs  sector, and one of them is the Planck constant. This is why the $X,Y$  vector boson masses and the Susyon mass are predicted, along with many other bosonic masses.  The quark matter CKM matrices,  and their leptonic counterparts, need the usual number of parameters that are familiar from the Standard Model without SUSY.
\item
The theory is highly constrained by the new Master Equation, which arises from the usual SU(5) SUSY GUT theory, coupled to supergravity, through simple \ET s.  Then loop corrections are governed by the BRST cohomology as in \ci{dixonnucphys}.
\item
The Master Equation means that the theory is valid for all momenta, and is renormalized in the way that was worked out for gauge theories before the advent of the idea of `effective theories', which have no Master Equation, and which are valid only for low momenta, and which lose control of all symmetries as a result.  Note that although the theory is not renormalizable, the Master Equation still controls its symmetry, no matter how many new terms arise. 

\item
In \fS, there is a necessary and simple form for a WIMP, and it might  account for dark matter. Can it be understood in terms of cosmological constraints? Does an extremely massive particle of this kind make any kind of a detectable track as it passes through the earth, on the rare occasions when it does so? 
Does such a particle have any kind of effect like that of a tiny black hole?  What is its effect in terms of quantized gravity? 
\item
This theory ought to be derivable from the superstring somehow, but I do not  see how.  
\item
What happens, for example, for theories based on SO(10) here? Or SO(32)? 
\item
Presumably it is possible to 
understand the weak angle using renormalization group arguments, as was done for the original SU(5) GUT theory.  
\item
It appears that this theory is chiral anomaly free \ci{ross,GUT}.
\item
There are a number of one loop issues.  One important issue is what happens to the cosmological constant at one loop in this theory.  
\een

 \refstepcounter{orange}
{\bf \theorange}.\;
{\bf  Perhaps there is a lesson to be learned here from the recent revolution in Quantum Mechanics, which has now given rise to the entire field of 
Quantum Information}.  The relatively obscure paper by EPR   seemed an academic curiosity until Bell  
showed that it had profound consequences for local realism, and this was then tested by experiments such as those by Aspect.  The history and experiments here and these counter-intuitive issues are dealt with nicely  in modern texts such as  \ci{lebellac}.  What we can perhaps learn here is that our notions, and even the instinctive and extremely reasonable `local reality' notions of Einstein, may need revision from experiment and futher thought.

 \refstepcounter{orange}
{\bf \theorange}.\;
{\bf \fS\ suggests that our intuitive notions of invariance may also be up for some revision.} It seems natural, and indeed obvious, to think that the Zinn sources are merely a convenience to formulate the BRST identities and the Master Equation.  If \fS\ has any validity, it looks like the Zinn sources may  have a more dynamic role, which goes beyond our ideas about symmetry being a reflection of the Noether type ideas that symmetry must be contained solely within the fields that we start with.    Without any doubt, \fS\ contains lots of fundamental problems that are not yet apparent, as do all attempts to understand these difficult questions.

 \begin{center}
Acknowledgments 
\end{center}

I thank Peter Scharbach for valuable remarks and insight.  I also thank Dylan Harries for help with writing the code in \ci{Dixon:2017eej}.

\tiny
\articlenumber\\
\today
\end{document}

%% file: Johnmacros25.tex
\newcommand{\fS}{Frozen SUSY}

\newcommand{\GUST}{{\rm SU(5) Grand Unified Supergravity Theory}}
\newcommand{\SGUST}{{\rm SU(5) Grand Unified Supergravity Theory with Suppressed SUSY}}

 \newcommand{\ET}{Exchange Transformation}

\newcommand{\sS}{Suppressed SUSY}

\newcommand{\pas}{\slash \hspace{-.5em} \pa}

\newcommand{\cM}{{\cal M}}

\newcommand{\cG}{{\cal G}}
\newcommand{\cK}{{\cal K}}

\newcommand{\cA}{{\cal A}}

\newcommand{\PB}{Master Equation}

\newcommand{\bt}{\begin{tabular}{c}}
\newcommand{\et}{\end{tabular}}

\newcommand{\eb}{\ee\be } 
 
\newcommand{\bmat}{\lt ( \begin{array} }
\newcommand{\emat}{  \end{array} \rt )}

\newcommand{\oH}{{\ov H}}

\newcommand{\oJ}{{\ov J}}

\newcommand{\ED}{\end{document}}

\newcommand{\oy}{{\ov \y}}

\newcommand{\oC}{{\ov C}}

\renewcommand{\a}{\alpha}	
\renewcommand{\b}{\beta}
\newcommand{\g}{\gamma}
\renewcommand{\d}{\delta}

\newcommand{\z}{\zeta}
\newcommand{\h}{\eta}
\newcommand{\q}{\theta}
\renewcommand{\k}{\kappa}
\newcommand{\lam}{\lambda}
\newcommand{\m}{\mu}
\newcommand{\n}{\nu}

\renewcommand{\r}{\rho}
\newcommand{\s}{\sigma}

\renewcommand{\c}{\chi}
\newcommand{\y}{\psi}
\newcommand{\w}{\omega}
\newcommand{\G}{\Gamma}

\renewcommand{\P}{\Pi}	
\renewcommand{\S}{\Sigma}

\newcommand{\la}{\label}
\newcommand{\ci}{\cite}

\newcommand{\ds}{\documentstyle}	
\newcommand{\fr}{\frac}

\newcommand{\na}{\nabla}
\newcommand{\pa}{\partial}
\newcommand{\ov}{\overline}
\newcommand{\be}{\begin{equation}}
\newcommand{\ee}{\end{equation}}
\newcommand{\ba}{\begin{array}} 
\newcommand{\ea}{\end{array}}
\newcommand{\bea}{\begin{eqnarray}}
\newcommand{\eea}{\end{eqnarray}}
\newcommand{\ra}{\rightarrow}
\newcommand{\Ra}{\Rightarrow}

\newcommand{\lt}{\left}
\newcommand{\rt}{\right}

\newcommand{\Tr}{{\rm	Tr}	}

\newcommand{\ben}{\begin{enumerate}}
\newcommand{\een}{\end{enumerate}}
\newcommand{\bitem}{\begin{itemize}}